# Associative Memory Impairments arising from Neurodegenerative Diseases and Traumatic Brain Injuries in a Hopfield Network Model

Melanie Weber*, Pedro D. Maia and J. Nathan Kutz

Department of Applied Mathematics, University of Washington, Seattle, WA 98195-3925, United States

* Corresponding author.

(Contact: melwe@uw.edu, pmaia@uw.edu, kutz@uw.edu)

## Abstract

Neurodegenerative diseases and traumatic brain injuries (TBI) are among the main causes of cognitive dysfunction in humans. Both manifestations exhibit the extensive presence of focal axonal swellings (FAS). FAS compromises the information encoded in spike trains, thus leading to potentially severe functional deficits. Complicating our understanding of the impact of FAS is our inability to access small scale injuries with non-invasive methods, the overall complexity of neuronal pathologies, and our limited knowledge of how networks process biological signals. Building on Hopfield's pioneering work, we extend a model for associative memory to account for FAS and its impact on memory encoding. We calibrate all FAS parameters from biophysical observations of their statistical distribution and size, providing a framework to simulate the effects of brain disorders on memory recall performance. A face recognition example is used to demonstrate and validate the functionality of the novel model. Our results link memory recall ability to observed FAS statistics, allowing for a description of different stages of brain disorders within neuronal networks. This provides a first theoretical model to bridge experimental observations of FAS in neurodegeneration and TBI with compromised memory recall, thus closing the large gap between theory and experiment on how biological signals are processed in damaged, high-dimensional functional networks. The work further lends new insight into positing diagnostic tools to measure cognitive deficits.

# 1 Introduction

Neurodegenerative diseases and traumatic brain injuries (TBI) are responsible for an overwhelming variety of functional deficits, both cognitive and behavioral, in animals and humans. Memory impairment, which is the focus of this work, is a particularly pernicious consequence for those affected. The pathophysiology of these induced brain disorders is usually complex, with key effects of the injuries occurring at small spatial scales that are currently inaccessible by non-invasive diagnostic techniques. Indeed, a hallmark feature of the injury/disease pathology is the abundance of diffuse, focal axonal swellings (FAS) which compromise spike train encodings in neuronal networks. Thus, there is a broad need to understand how neuronal pathologies which develop at a cellular level compromise the functionality of a network of neurons responsible for cognitive function. In this article we extend well-established computational models of associative memory, i.e. the Hopfield network model, to incorporate FAS pathologies implicated in many brain disorders, providing novel metrics for quantifying memory impairments and functional deficits. The work is the first of its kind to integrate extensive experimental findings regarding FAS with mathematical models of spike-train encoding, neuronal networks, and memory association.

Human memory is conjectured to work by forming associations [Gerstner et al(2014), Hopfield(1982), Squire et al(2007)]. Meaningful stimuli are usually encoded, stored, and later recalled by the brain in response to some cue for use in a given activity. This cognitive process improves the interpretation of subsequent stimuli that share common features with previously stored concepts [Hopfield(1982)]. What governs our memory recall abilities are coordinated exchanges of electrical signals between neurons in network structures. In fact, neurons constantly encode and transmit information in the form of spike trains along their axons [Gerstner et al(2014)]. Several pathological effects, most notably FAS, can jeopardize this critical electrical activity, making memory performance particularly sensitive to common brain disorders. Given its ubiquity across many neurodegenerative diseases and traumatic brain injury, understanding the role of FAS in altering spike train encodings is of paramount importance, particularly on a network level where cognitive functionality occurs. In this work, we incorporate the biophysical effects of FAS observed in neurodegenerative diseases and/or following TBI to network models that exhibit memory retrieval capabilities. Specifically, we integrate cable theory models with FAS to quantify the impact of swellings on spike train propagation [Maia, Kutz(2014a), Maia, Kutz(2014b), Maia et al(2015)] and integrate these findings with a Hopfield associative memory model.

Neuroscience has historically combined experimental observations with theoretical models, from the early neuronal doctrine and single-cell analysis [Cajal, Estructura(1888), Bullock et al(2005)] to more recent neuronal-network paradigms [McCulloch, Pitts(1943), Hebb(1949)]. Experimental advances in physiological measurements of neuronal ensembles continuously challenge and improve existing models, leading to larger and more complex simulations [Yuste(2015)]. Today, computational neuroscience is recognized as an important and accessible alternative to modeling the dynamical interplay between neurons, especially when biophysical recordings are limited and/or inaccessible. Remarkably, John Hopfield's pioneer model for associative memory, and its later variations, are still widely used today to describe the recall of previously stored mental patterns through the evolution of neuronal dynamics [Hopfield(1982), Hopfield(1984), Hopfield, Tank(1985), Hopfield(2015), Hu et al(2015), Nowicki, Siegel-

mann(2010)]. Our aim is to modify the Hopfield model of associative memory by including the deleterious effects of FAS. Due to its simplicity and popularity, we found the Hopfield network model to ideally suited for first investigations of established single cell models for memory impairment [Maia, Kutz(2014a), Maia, Kutz(2014b), Maia et al(2015)] at the level of neuronal dynamics. The rich body of knowledge that has been developed for the Hopfield model serves as an ideal basis for this study and allows for an embedding of our results in existing theories [Anafi, Bates(2010), Menezes, Monteiro(2011), Ruppin, Reggia(2011)].

Overwhelming experimental evidence suggests that FAS is the hallmark manifestation of injuries on neuronal networks. Moreover, there now exists a wealth of experiments characterizing the statistical distribution of FAS as a function of injury level, including size and frequency of swellings [Coleman(2005)]. Statistics can even be collected for specific neurodegenerative diseases such as Alzheimer's [Krstic, Knuesel(2012)], Parkinson [Galvin et al(1999)] or Multiple Sclerosis (MS) [Hauser et al(2006)], where the swellings occur as a consequence of complicated biophysical/biochemical deterioration of neurons. Thus, biophysically plausible models can be posited that make use of the best available data regarding neurodegeneration and TBI.

Understanding the role of FAS is of critical importance for comprehending cognitive deficits and memory impairment. Neurodegeneration via Alzheimer's, Parkinson's or MS is estimated to affect a large portion of adults. Indeed, Alzheimer's alone has recently be estimated to be the third leading cause of death, just behind heart disease and cancer, for older people. Likewise, TBI is responsible for millions of hospitalizations worldwide every year and is the leading cause of death among youngsters [Faul et al(2010), Jorge et al(2012), Xiong et al(2013)]. TBI is usually caused by external forces applied to the head as occurs often in contact sports. In fact, it has been reported [Fainaru, Fainaru(2013)] that $10 - 20\%$ of professional football players suffer from TBI. The external forces damage neuronal axons in the brain resulting in FAS that modifies the shape of the axons, which in turn, distorts the transmitted pulses leading to a potentially significant loss of information [Maia, Kutz(2014b), Tang-Schomer et al(2010), Tang-Schomer et al(2012)].

## 2 Results

In this article, we introduce a biophysically realistic variation of Hopfield's model to incorporate a single-cell neuron model for FAS [Maia, Kutz(2014a)]. The FAS modeling is based upon state-of-the-art statistical estimates of size and distribution of FAS observed in experiments. We demonstrate the impact of FAS with intuitive examples from face recognition. From a translational viewpoint, our models give clear metrics for evaluating memory deficits as a function of injury level, thus suggesting how memory impairment metrics may allow one to infer network damage levels from memory recall tasks.

## 2.1 Face recognition on associative-memory networks

We calibrate a Hopfield-like neuronal network [Hopfield(1982), Hopfield(1984)] to perform face recognition tasks (see Fig. 1). Our sample memory space consists of five human facial images (1044 × 1341 pixels), modified from the *MIT faces database* [Weyrauch et al(2004)] [1]. For the second experiment (Fig. 2) we choose the three most correlated images to test for confusion between closely related memories. In our setting, neurons dynamically alternate between multiple firing-rate activity states as opposed to the binary formulation in the original standard Hopfield model [Hopfield(1982)]. This more realistic higher-dimensional feature space allows for the encoding and storing of more complex memories in the network.

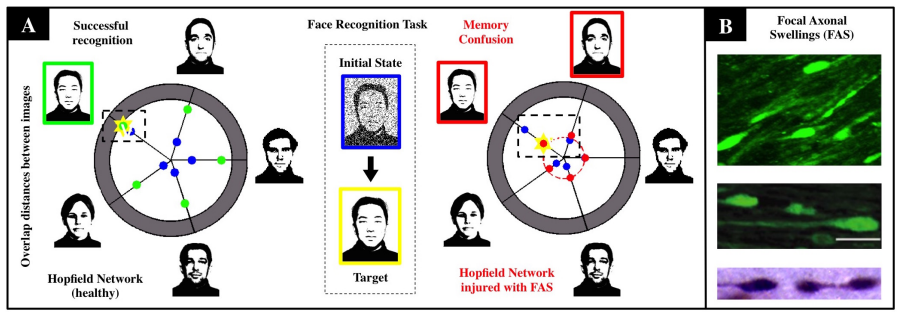

**Fig 1.** Face recognition task under the influence of neurodegenerative diseases and traumatic brain injury. **A:** For a noisy input with 80% overlap (i.e. 20% initial noise), the healthy system (left) achieves an overlap of 90% with the correct face (marked yellow). The chart presents the initial (blue) and final (green) Hamming distances between the current network state and the patterns corresponding to the respective facial images. When the recognition task is performed with an injured system (right), we observe a severe decrease in accuracy (overlap significantly smaller than in the healthy network) and confusion between two facial images. Both are characteristic injurious effects of memory impairments arising from FAS and are quantified with a recognition score in our study. **B:** Images of FAS – highlighting their morphometric features– adapted from experimental works used to calibrate our model [Wang et al(2011), Dikranian et al(2008)].

The weighting of neuronal connections captures key aspects of the network architecture, including how memories are encoded. They are initially calibrated to encode the desired set of memorized concepts as fixed points of the system. We add Brownian noise fluctuations to the dynamics as a proxy for natural stochastic fluctuations. As a result, when a partial or noisy version of one of the memory concepts is presented, the system converges to the closest fixed point matching the input to the closest (least-square fit) memory. Our healthy network demonstrated in Fig. 1 achieved a recall accuracy of approximately 90% and distinguished perfectly between all images in our sample set. Injuries due to FAS deteriorate network performance by misclassifying given input stimulus, thus creating potential confusion in memory association.

---

[1] Credit is hereby given to the Massachusetts Institute of Technology and to the Center for Biological and Computational Learning for providing the database of facial images. http://cbcl.mit.edu/software-datasets/heisele/facerecognition-database.html

## 2.2 Modeling dynamical impacts of FAS.

The presence of FAS is known to distort neuronal spike-train dynamics, but precise electrophysiological recordings of pre- and post-FAS spike train dynamics are still unavailable. The recent theoretical models of Maia and Kutz [Maia, Kutz(2014a),Maia, Kutz(2014b),Maia et al(2015)], however, provide important estimates of spike train deterioration due to FAS. Using this model, we can characterize the effect of injury on the firing rate activity through a response function. The state variable of a Hopfield node, $S_i$, is equivalent to the firing rate of that node. The collection of all nodes is denoted by the vector $\mathbf{S}$. Injuries can then be characterized by the transfer function

$$\tilde{\mathbf{S}} = F(\mathbf{S}, \beta). \qquad (1)$$

where $\tilde{\mathbf{S}}$ is the effective firing rate (state) after the FAS, $\mathbf{S}$ is the firing rate (state) before the FAS, and $\beta$ is a parameter vector indicating one of three injury types applied to individual nodes of the network [Maia, Kutz(2014a),Maia, Kutz(2014b)]. The function $F(\cdot)$ maps the pre-injury to post-injury firing rates using biophysically calibrated statistical distributions of injury in both frequency and size. If no FAS occurs to a given axon, then its state is unaffected ($\beta_0$). This occurs with probability $1-p$ where $p$ is the probability of injury and what we term *injury level*, i.e. larger $p$ implies more injury. For those neurons with axonal swellings, the following manifestation of spike train deformations have been observed: unimpaired transmission ($\beta_1$), filtered firing rates ($\beta_2$), spike reflection ($\beta_3$), or blockage ($\beta_4$). The injury type is dependent upon the geometry of the swelling, with blockage being the most severe injury type. From biophysical data collected on injury statistics [Dikranian et al(2008),Wang et al(2011)], both in swelling size and frequency, we assign a prescribed percentage of each type of injury ($\beta_j$) to the network using calibrated simulations of spike propagation dynamics [Maia, Kutz(2014a),Maia, Kutz(2014b)]. For a blockage injury ($\beta_4$), no signal passes the swelling so the effective firing rate of the neuron goes to zero. Thus, $\tilde{\mathbf{S}} = 0$ which prevents the neuron from adapting to the collective dynamics. Filtering injuries were taken to decrease the firing rate, with higher firing rates having a stronger chance of decreasing due to pile-up effects in the spike train [Maia, Kutz(2014b)]. Reflection of spike trains effectively filters the firing rate of an axon by a factor of two so that $\tilde{\mathbf{S}} = 0.5\mathbf{S}$. This is due to the fact that the reflected spike annihilates an oncoming spike [Maia, Kutz(2014a)]. Overall then, the method for producing the filtering function $F(\cdot)$ uses the most advanced experimental findings to date with recent computational studies of spike train propagation through FAS [Maia, Kutz(2014a), Maia, Kutz(2014b)]. See the Materials and Methods section for details on how we compute the statistical distribution of injury types that are parametrized by the parameter $\beta_j$.

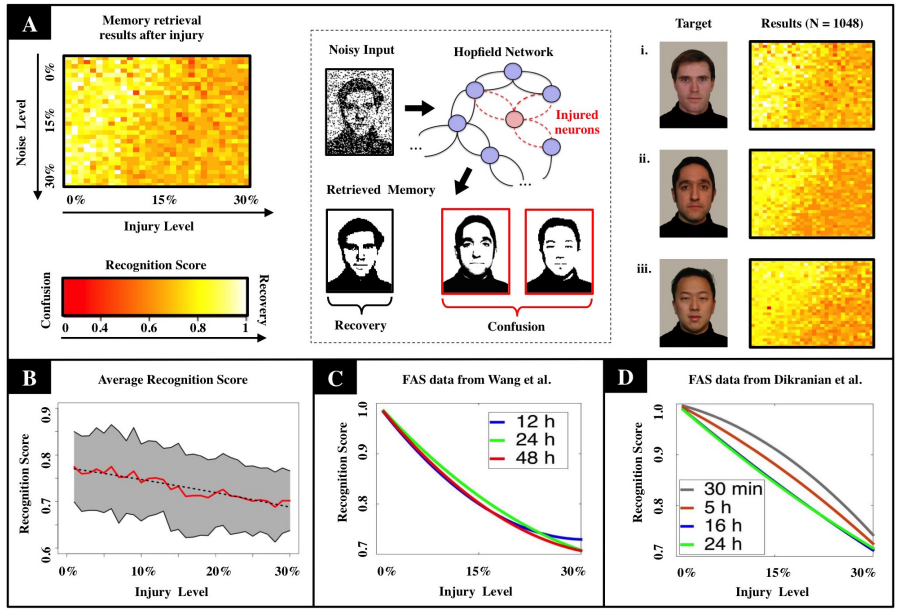

**Fig 2.** Statistical evaluation of computational results. **A:** Effects of FAS on associative memory for a face recognition task. We measure recognition ability and accuracy for a sample set of three facial images [Weyrauch et al(2004)] over a given range of parameters (injury: $0-30\%$, initial noise: $0-30\%$). As the degree of injury increases, the noise handling ability of the system drops severely: The coloring of the heat maps change from yellow (significant recognition) in the upper left (small injury and initial noise) to dark orange and red (confusion) in the lower right corner (high injury and initial noise). We observe confusion and a decline in recall accuracy for every image in the set of samples. Regression over all N=1080 normalized data points indicates an exponentially declining relation between recognition scores and injury level as shown in the diagram. **B:** Mean recognition scores are displayed in red, standard deviations are shaded in gray. The dashed line indicates the recognition function obtained by linear regression. **C,D:** We tested our model in a biological realistic setting using experimental results from TBI studies by Wang et al. [Wang et al(2011)] and Dikranian et al. [Dikranian et al(2008)]. Our results show a modest, but unstable increase in memory performance for the adult brain during the first 24 hours after the injury (**C**, [Wang et al(2011)]), that drops again to the initial level during the following day. In the infant's mouse brain, we observe a steadily decreasing performance in the first 24 hours (**D**, [Dikranian et al(2008)]). The observations match the long recovering times that are commonly observed in patients suffering from TBI.
6

The presence of a significant number of blocked neurons in the Hopfield network creates blurring in the reconstructed concepts (faces) and therefore decreases the accuracy of the recalled information. Both filtering and reflection lead to confusion of correct states with their neighboring ones (in a Hamming distance sense). Ultimately, this decreases the ability of the network to perform denoising tasks. In most cases of FAS, all three mechanisms occur in simultaneously, depending on the type and time development of the injury [Maia, Kutz(2014a)].

## 2.3 Memory impairments and confusion of concepts

We examine the memory-recall performance of neuronal networks as a function of (a) distributions of FAS, (b) time-lapse from injury, and (c) percentage $p$ of injured neurons. Our recognition score (see Fig. 6) quantifies the network's accuracy and ability to differentiate between previously stored concepts from noisy cues. Figure 2(a) shows recognition scores for the most similar triplet of faces in our set.
In our first numerical experiment we considered FAS distributed according to results by Wang et al. [Wang et al(2011)] for TBI injuries after 12 hours. We varied the amount of noise $(0-30\%)$ and degree of injury $(p = 0-30\%)$ for each of the facial images in our triplet, allowing for cross-comparison of recognition scores and memory performance with respect to the choice of parameters (see Fig. 2). An interesting consequence of injury is that the network can produce, and often does for elevated injuries, erroneous associations of memories. Specifically, it *confuses* the concepts in the Hopfield network. Based on simulations of an uninjured network with noise, it could easily be conjectured that the fixed points associated with a given memory would *disappear*, with the system's dynamics preventing it from converging to the correct pattern. On the contrary, we observe that the noise fluctuations cause the dynamics to converge to an erroneous fixed point under the influence of FAS in the injured system. For modest amounts of initial noise $(15 - 30\%)$, the affected system confuses the stored patterns and looses the ability to separate them properly. Confusion of concepts is especially pronounced when blockage and filtering account for the majority of FAS effects.
Focal axonal swellings decrease the system's ability to handle noise and recall a previously stored pattern from partial information. Our simulations allowed us to quantify the *rate* at which the system's average recognition score $R$ decrease as a function of injury strength $p$:

$$R(p) = A - Be^p \qquad (2)$$

The functional form $R(p)$ is produced from a linear regression over $N = 1080$ normalized data points obtained from computational experiments. We examined the development of FAS from two different experimental data sets: in an adult rat brain 12, 24 and 48 hours after TBI (Fig. 2c, [Wang et al(2011)]) and in an infant mice brain 30 min, 5, 16 and 24 hours after TBI (Fig. 2d, [Dikranian et al(2008)]). The values for the constants in eq. (2) are given by $[A = 1, B = 0.24]$ and $[A = 1.21, B = 0.22]$ respectively.

We also evaluate recognition scores on injured networks as a function of time. Our results hint at a modest, but unstable increase in memory performance for the adult brain during the first 48 hours after the injury, and a performance decrease for the infant brain in the first 24 hours (see Fig. 2). This is consistent with the relatively long recovering times observed in patients suffering from TBI.

## 3 Discussion

We consider memory impairments caused by Focal Axonal Swellings in a well-established Hopfield computational model for associate memory. The FAS are a hallmark feature of traumatic brain injuries and neurodegenerative diseases such as Alzheimer's, Parkinson's and MS. With this model, we can characterize a neural network's ability to handle noise and perform recognition tasks at different levels of injury. Our computational model reproduces symptoms commonly observed in patients suffering from brain disorders [Knutson(2012)]. The injured network-model for the task of face recognition loses accuracy in recalling facial images and confuses faces with similar features. From a Hopfield network viewpoint, less accuracy means that fixed points encoding facial images loose their stability. At higher injury levels, the network's dynamic is driven away from the correct face encoding, settling closer to an unstable fixed point associated with another facial pattern. Instability of the fixed points results from a large percentage of neurons manifesting FAS, leading ultimately to confusion between previously stored images.

Confusion of concepts has a higher impact on the overall performance than the loss of accuracy. It severely affects the *content* and *context* of a stored concept. Such impairments are commonly observed in patients suffering from brain disorders [Coleman(2005)], especially in advanced stages of Alzheimer's diseases [Selkoe(2001),McKhann et al(1984)] and severe traumatic brain injuries [Cole, Bailie(2016),McGee(2016)].

To test our injury protocols in a biologically realistic setting, we calibrate our simulations with experimental FAS data from Wang et al. [Wang et al(2011)] and Dikranian et al. [Dikranian et al(2008)]. Both studies perform traumatic brain injuries to their model organism: [Dikranian et al(2008)] studied FAS in the cortex of infant mice, whereas [Wang et al(2011)] investigated swellings in the optic nerve of adult rat. Although [Dikranian et al(2008)] provides detailed morphometric data for injured axons in the infant mice, their functional assessments are still far from interpretable studies of memory development in human children [Mcallister(2004)]. TBI experiments in adult rats [Xiong et al(2013)] report functional impairments more analogous to human patients – like deficits in tasks involving context memory [Schacter(1987)], conditional associative learn-ing [Petrides(1985)], planning [Shallice, Evans(1978)] and other cognitive tasks [McDowell et al(1997)]. Adult rats, however, may exhibit memory deficits after mild TBI even without many signs of axonal injury [Lyeth(1990)]. In fact, recent studies demonstrate that catecholamines play a central role in the neurochemical activation and regulation of working memory [Mcallister(2004)] and such effects were not incorporated in the model. Thus, axonal structural damage may be sufficient but not necessary for the production of neurological and cognitive symptoms associated with TBI. This will be considered in future studies.

There is much room for improvement in our injury model, especially given the variety of effects of different types of brain disorders. Although FAS are universal pathological features, our results should be regarded as a foundational mathematical framework to study different disorders in neuronal networks. We hope to continue to develop computational models capable of describing a wide variety of cognitive deficits. Here, we use noise-handling as a proxy for cognition and describe functional deficits in terms of a few key model parameters. Our ongoing work includes the study of convergence and stability in the extended associate memory Hopfield model and the impact of memory impairments on these aspects.

More importantly, this framework opens up opportunities for novel diagnostics: one could use a patient's inability to distinguish known visual impressions as an indirect proxy for cellular pathologies currently inaccessible to non-invasive techniques. There is a broad need in the neurological and biomedical communities to understand the development, propagation, and functional implications of brain degeneration. As a consequence, medical imaging evolved dramatically with our ability to extract more information from highly under-sampled noisy data. Advances in Diffusion Tensor Imaging (DTI), functional MRI, accelerated MR imaging, and brain morphometry allow us, for the first time, to unveil network characteristics that distinguish healthy and pathological brains for a number of brain disorders. In fact, data-driven methods are key to graph-cut neuroimaging software, high angular resolution diffusion imaging (HARDI), and network-level analysis of cortical thickness from structural MRI. These methods, when combined, enable whole-brain network reconstruction and are revolutionizing our understanding of incurable neurodegenerative diseases like Alzheimer's, Epilepsy, Autism, Schizophrenia, Stroke, and others. Such diseases, however, pose two fundamentally difficult challenges for functional diagnostics; changes in structural connectivity do not translate naively to dynamical neuroactivity. In fact, understanding information processing and network functionality is already a daunting task even in healthy brains. Secondly, there are pathological developments in spatial scales inaccessible to non-invasive techniques. At a cellular level, for instance, the presence of Focal Axonal Swellings (FAS) is a critical biomarker for neurodegeneration and traumatic brain injuries. Differences in their geometrical ultrastructure can affect neuronal signaling and degrade the information encoded in spike trains. Our recent estimates highlight potential axonal trouble spots and provide a window of opportunity for theoretical and computational exploration of their collective effects in impaired neuronal networks.

Theoretical and computational models may – with all their limitations – link injured neuronal networks to observable psychophysical deficits. Although much remains to be explored, the associate memory model based upon the classic work of Hopfield subject to FAS represents the first fundamental, biophysically-inspired theoretical model addressing cognitive deficits due to injury.

## 4 Methods and Materials

### 4.1 Hopfield network models for associative memory

J. Hopfield made seminal contributions to the study of collective properties that emerge on systems of equivalent components (or neurons). He developed a model to describe content-addressable memory in an appropriate phase space for neuronal networks. The model incorporated aspects of neurobiology and its underlying neuronal circuitry. Within this framework, he was able to study properties such as familiarity recognition, categorization, error correction, and time sequence retention. Our computational model is based on J. Hopfield's original publications [Hopfield(1982), Hopfield(1984), Hopfield, Tank(1985)] and more recent extensions [Gerstner et al(2014), Benna, Fusi(2015)].

#### 4.1.1 Standard Hopfield Model

In Hopfield's original model, a neuronal network is composed of $N$ neurons that attend binary states $S_i \in \{-1, 1\}$. The connections between neurons are responsible for information transference and processing in the network. They are represented by weights $w_{ij}$ (linking neurons $i$ and $j$), and stored in a connectivity matrix $W = (w_{ij})$. In this setting, the neuronal states evolve in time according to [Hopfield(1982), Hopfield(1984), Hopfield, Tank(1985)]

$$dS_i(t) = \sum_j w_{ij} \cdot g(S_j(t)) dt \qquad (3)$$

where the gain function $g$ is given by

$$g(x) = \begin{cases} 1, & x \geq 0 \\ -1, & x < 0 \end{cases} . \qquad (4)$$

The most important property of the model is the ability to encode memories as fixed points of the system. When a noisy input is presented, it converges to the closest fixed point (closest known concept) in a process commonly referred as *memory association*.

#### 4.1.2 Extended Hopfield Model

In our extended Hopfield model, neurons may achieve multiple discrete states [Gerstner et al(2014), Benna, Fusi(2015)]

$$S_i \in \{0, 1, ..., s-1, s\}.$$

The dynamical evolution of the system is also governed by a more sophisticated equation:

$$dS_i(t) = \underbrace{-\tau^{-1} \cdot S_i(t)dt}_{\text{self-dynamics}} + \underbrace{I_i(t)dt}_{\text{external input}} + \sum_j \underbrace{w_{ij} \cdot g(S_j(t))dt}_{\text{input from other neurons}} + \underbrace{\mu \cdot dB_i}_{\text{noise}}, \quad (5)$$

with sigmoid gain function $g$ given by

$$g(x) = 0.5(1 + \tanh(\beta x)). \quad (6)$$

The constant $\tau$ gives the time-scale of the dynamics. Direct inputs for neuron $i$ (e.g. external stimuli) are represented by $I_i(t)$. The term $B_i$ corresponds to a Wiener Process with intensity $\mu$, and is a proxy for stochastic fluctuations in the firing rates. The (continuous) states are ultimately rounded to the nearest discrete state by a scaling function

$$m(t) = \max_{1 \leq i \leq N} |S_i(t)| \quad (7)$$

$$\hat{S}_i(t) = \left[\frac{s \cdot S_i(t)}{m(t)}\right] \quad (8)$$

The resulting stochastic differential equation takes the following form when discretized:

$$S_{i+1} = S_i + \Delta t\, f_1(S_i) + f_2(S_i)\, dB_i \quad (9)$$

with $f_1(S_i) = -\tau^{-1} S_i + \sum_j w_{ij} g(S_j) + I_i$, $f_2(S_i) = \mu S_i$

and Brownian increment $dB_i = B_i(t + \Delta t) - B_i(t)$. We solved the system numerically using the *Euler-Maruyama Method* [Higham(2001)] and made all our codes available.

As noted earlier, FAS may alter the activity state of an injured neuron. Thus, a broader class of states is required to study effects of more sophisticated mechanisms. Whereas blockage of spike trains can be described in binary networks, more complex feature discriminations like reflection and filtering, that transfer partial or altered information, require a multi-state setting. For this, we converted our sample set of facial images [Weyrauch et al(2004)] from originally colored to nine distinct shades of gray ($s = 8$) in addition to the (standard) binary black-white states. With this extension, the computational network can also discriminate images of human faces that naturally share several features (high overlap).

## 4.2 Focal Axonal Swellings following brain injuries

Focal Axonal Swellings are an ubiquitous pathological feature to several brain disorders [Xiong et al(2013), Coleman(2005), Tang-Schomer et al(2010), Tang-Schomer et al(2012), McArthur et al(2004), Bayly(2006b)]. Recent theoretical results suggest how morphological parameters affect the propagation of spike trains and thus, disrupt functional and coordinated network activity. In this section we provide details on how we introduce such effects in our Hopfield-networks and detail the experimental work used to calibrate the parameters of our model.

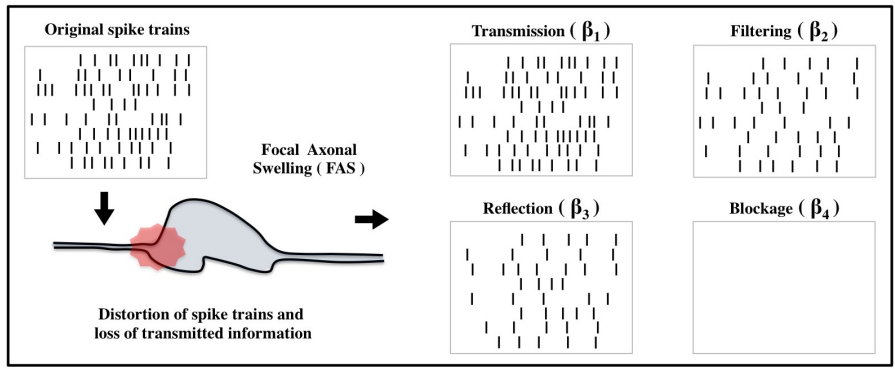

**Fig 3.** Schematics for injurious effects of Focal Axonal Swellings (FAS) to spike trains according to the theoretical framework of Maia and Kutz [Maia, Kutz(2014a), Maia, Kutz(2014b), Maia et al(2015)]. Spike trains may be distorted in four qualitatively distinct regimes depending on geometrical parameters associated to FAS: transmission ($\beta_1$), filtering ($\beta_2$), reflection ($\beta_3$) or blockage ($\beta_4$). See references and text for more details. Injuries can then be characterized by the transfer function $\tilde{\mathbf{S}} = F(\mathbf{S}, \beta)$ where $\tilde{\mathbf{S}}$ is the effective firing rate (state) after the FAS, $\mathbf{S}$ is the firing rate (state) before. Such distortions are incorporated to injured neurons in our Hopfield network simulations in conjunction with experimental morphometric results from Wang et al. [Wang et al(2011)] and Dikranian et al. [Dikranian et al(2008)].

### 4.2.1 Theoretical framework for FAS effects

Maia and Kutz developed in a series of papers a theoretical framework for characterizing the anomalous effects of FAS to spike propagation [Maia, Kutz(2014a), Maia, Kutz(2014b), Maia et al(2015)]. We review their main results and explain how to add such pathologies (or their proxies) into account for the firing-rate dynamics of neuronal networks (see the schematics in Fig. 3).

The authors distinguish axonal enlargements that lead to minor changes in propagation ($\beta_1$) from those that result in critical phenomena such as collisions, reflections or blockage of traveling spikes ($\beta_2, \beta_3$ and $\beta_4$). They use three geometrical parameters ($\delta_B, \delta_T, \delta_A$) to model a prototypical shaft enlargement and characterize all possible propagation regimes in an unmyelinated action potential model. The regimes can be distinguished by evaluating a (simple) function of the FAS geometrical parameters inferred through numerical simu-lations. They suggest that evaluating this function along axon segments can help detect regions most susceptible to (i) transmission failure due to perturbations, (ii) structural plasticity, (iii) critical swellings caused by brain traumas and/or (iv) neurological disorders associated with the break down of spike train propagation.

Swellings typically delete spikes by a mechanism called *filtering* ($\beta_2$), when a first spike changes its profile at the axonal enlargement region and a close second spike interacts with its refractory period. As a consequence, the second spike is deleted in a mechanism of the so-called *pile-up collision* (see [Maia, Kutz(2014b)] for details). Distorted spike trains do not match their corresponding original firing rates (as illustrated in Fig. 3). Instead, they are confused with lower rates, which decrease the system's overall denoising abilities. We simulate the harmful effects of filtering by implementing a statistical version of the confusion matrix from the same source, that in simple terms, evaluates the probability that state $i$ gets confused as state $j$ due to the FAS.

A less frequent mechanism of spike deletion is *reflection* ($\beta_3$). There, a traveling pulse is divided into two pulses when it reaches the FAS: one propagating forwards and the other propagating backwards. The backward pulse will collide with the next spike and have them both deleted. Thus, only a fraction of the original encoded information is ultimately transmitted by the spike train. We add this effect in our neuronal network by halving the firing-rate of an injured neuron in this regime.

Figure 3 finally illustrates the *blockage* of spikes ($\beta_4$) that occur typically in regions of more dramatic axonal enlargement. In this scenario, no information is transmitted through the damaged axon and the neuron cannot adapt and play its role in the desired collective dynamics: it remains in its initial state. In our neuronal network model, a significant amount of neurons in the blockage regime causes blurs in the reconstructed concepts (memories) and therefore decreases the accuracy of the recalled information. We modeled this mechanism by introducing non-adapting neurons into the network, that keep their (possibly noise-affected) initial state over time.

Traumatic Brain Injuries and neurodegenerative diseases induce FAS with tremendous variety of shapes and, consequently, with different functional deficits regarding spike train propagation. Thus, we consider a *distribution* of different FAS mechanism where fractions of neurons are affected by blockage, reflection and filtering (confusion) respectively.

### 4.2.2 TBI/FAS data from adult rats (Wang et al. [Wang et al(2011)])

We use the data from Wang et al. [Wang et al(2011)] to calibrate the distributions of FAS following TBI in our simulations. In their study, the authors induce axonal injury in the optic nerve of a rat via central fluid percussion injury. They follow the progression of proximal and distal axonal injuries with biological markers for swellings. Transgenic animal models were used to better highlight damaged axonal segments with immunocytochemical markers. Through the use of qualitative and quantitative imaging approaches, they confirm and significantly extend previous data supporting the pathogenesis of traumatic axonal injury. Importantly, they provide unprecedented insight into this complex pathology, demonstrating post-TBI axonal swelling and disconnection. Within this setup, any potential axonal change induced by the traumatic event could be easily and consistently visualized via routine fluorescence and/or confocal microscopy.

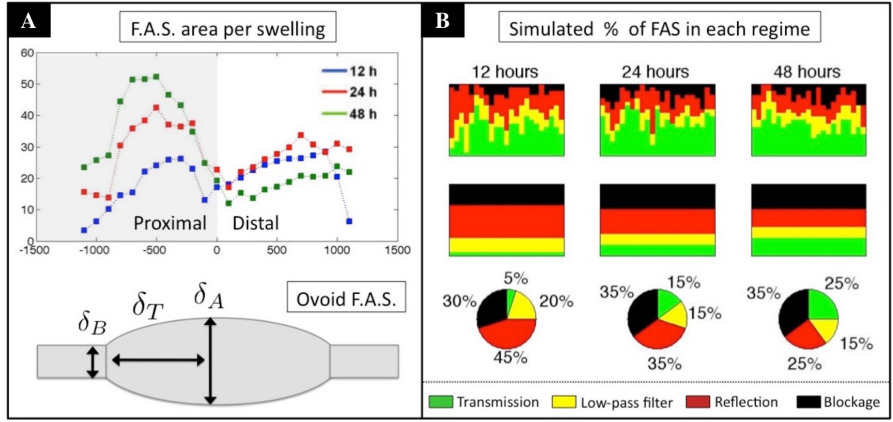

**Fig 4.** Distribution of functional impairments in injured axons following TBI experiments from Wang et al. [Wang et al(2011)]. Panel **A**: we generate ovoid/spheroid FAS with areas compatible with the experimental distribution. The geometrical parameters of the FAS define the spike propagation regime. Panel **B**: we generate 12 FAS (column) for each injured axon (row) and order them from worst to best case scenario (upper "flags"). We assume that the worse FAS within an injured axon dominates the others, and classify the entire axon within that category (intermediary "flags"). This leads to the (bottom) pie-charts of impairments for an injured neuronal population. See text for more details.

In this work we are especially interested in the distributions of shape, number and area of FAS from Wang et al. [Wang et al(2011)]. The number and area of FAS vary with time and spatial localization along the optic nerve. This allowed us to reconstruct distributions of total area per swelling and their functional deficits (blockage, reflection and filtering). See Fig. 4 for details. Wang et al. [Wang et al(2011)] describe a "spheroid" as a common type of FAS shape. We model this particular type of FAS with ovoids (see Fig. 4A) that include the three geometrical parameters $(\delta_B, \delta_T, \delta_A)$. The total area of each generated FAS is drawn from the distribution of area per swelling depicted in Fig. 4A (within one standard deviation). The authors divided the optic nerve into 12 spatial grids: distal (D1-D6) and proximal (P1-P6). We generate one swelling per grid for each injured axon and order them from worst to best case scenario: Fig. 4B (upper "flag-charts") should be interpreted as a family of 20 injured axons (rows) with 12 FAS each (columns). Each FAS has a functional deficit (transmission, filtering, reflection or blockage) according to its geometrical parameters. Finally, we assume that the worst swelling within an axon (row) dominates and classifies the entire axon within that category (Fig. 4B, intermediate "flag-charts"). This lead to the pie-charts indicating the fraction of axons displaying transmission (green), filtering (yellow), reflection (red) and blockage (black).

There are several drawbacks with this methodology, but we believe that our distributions are biophysically reasonable and compatible with available data. Better results could be obtained if the authors used the recently diagnostic tools developed in [Maia et al(2015)].

### 4.2.3 TBI/FAS data from infant mice (Dikranian et al. [Dikranian et al(2008)])

Traumatic Brain Injuries (TBI) are responsible each year for millions of children hospitalizations [McArthur et al(2004)]. Infant brains (0-2 years) are highly susceptible to trauma [Schneier et al(2006)], and injuries sustained during this critical development period leads to profound neuronal and axonal degeneration [Gleckman et al(1999), Ewing-Cobbs(2000), Geddes et al(2001)]. K. Dikranian et al. [Dikranian et al(2008)] examined analogous injuries in the infant P7 mouse due to its physiological similarities with human children, inflicting controlled impact to the skull with an electromagnetic device based on a moving coil actuator. Animals were sacrificed at 30 min, 5h, 16h and 24h post-injury and well-established methodologies [Bayly(2006b)] were used to characterize the spatiotemporal pattern of axonal degeneration and cell death in the cingulum/external capsule region ipsilateral to the site of trauma. The Cingulate Cortex plays a critical role in memory performance [Vann et al(2012), Malin, McGaugh(2006), Weible et al(2012)], and they suggest that early degeneration there may disconnect cortical and thalamic neurons, leading to their apoptotic death.

Dikranian's morphometric analysis [Dikranian et al(2008)] – that included detailed swelling diameter measurements – allowed us to realistically incorporate injury effects in our Hopfield network simulations. We are especially interested in their distributions of shape, number and area of FAS. Notice that they provided distribution for the diameters of the spheroids, which allowed us to generate FAS in a much more direct way. The corresponding "flag-charts" and pie-charts for their distributions are depicted in Fig. 5. We believe these distributions to be more relevant than the ones obtained using the data from Wang et al. [Wang et al(2011)] since the cingulum is recognized as a region that plays a fundamental role in memory association. Thus, damage in this area can be also physiologically linked to memory impairments.

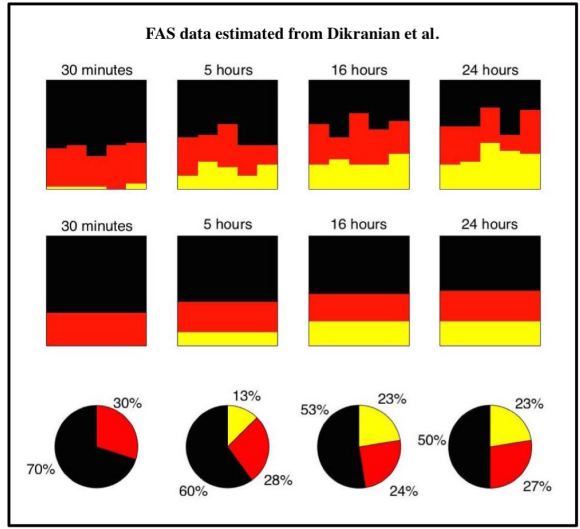

**Fig 5.** Distribution of functional impairments in injured axons following TBI experiments from Dikranian et al. [Dikranian et al(2008)]. We generate ovoid/spheroid FAS following the reported experimental distribution of FAS diameters. The geometrical parameters of the FAS define the spike propagation regime. We generate 5 FAS (column) for each one of the 40 injured axons (row) and order them from worst to best-case scenario (upper "flags"). We assume that the worse FAS within an injured axon dominates the others, and classify the entire axon within that category (intermediary "flags"). This leads to the (bottom) pie-charts of impairments for an injured neuronal population. See text for more details.

### 4.3 Implementation of memory storage.

To simulate a face recognition task, the set of memories has to be *learned* by the network. For this, we encode them in the weights of the neuronal connections as specified by the *weight matrix* of the network:

We consider a system of weighted neurons. The strength $w_{ij}$ of the connection between neuron $i$ and neuron $j$, described by the weight of the respective edge, characterizes the information transfer from i to j. Stored in the connectivity matrix $W = (w_{ij})_{1 \leq i,j \leq N}$, they characterize the network's dynamics and encode the set of known concepts corresponding to the system's fixed points.

The weight matrix is constructed from the training set of memories represented

as network states:

$$C := \left\{ \begin{matrix} \vdots & \vdots & \ldots & \vdots \\ \text{face 1} & \text{face 2} & \ldots & \text{face M} \\ \vdots & \vdots & \ldots & \vdots \end{matrix} \right\} \rightarrow W = C^T C \qquad (10)$$

The theoretical storage capacity of a (Standard) Hopfield network of size $N$ is $0.14N$ random patterns. In this study, we use a much smaller set of memories, respectively five and three. This is due to the fact, that we store highly correlated facial images as opposed to random patterns. They have a pairwise overlap of 60% due to the structural similarity of faces. The high correlation pf the memories significantly decreases the storage capacity and therefore requires the choice of a small set of memories. We choose a setting with highly correlated memories to demonstrate the effects of memory confusion arising from FAS as described earlier.

### 4.4 Recognition score for network performance

We developed a recognition score that measures recognition abilities with respect to significance and accuracy in recalling previously stored memory patterns (see Fig. 6).

We assume the existence of an *ideal observer* (cf. Benna and Fusi [Benna, Fusi(2015)]), that knows the whole set of memories and the original pattern underlying the current noisy input. Our recognition score takes the place of this observer by evaluating the current network state against all memorized patterns. In what follows, we describe the computational steps of the recognition algorithm:

(i) *Overlap:* We determine the overlap between the current network state and the set of stored memories by calculating the respective overlap $m^\mu \in \{0, 1\}$ of individual neuronal states:

$$m^\mu = \frac{1}{N} \sum_j \delta_{|j - \hat{j}| < 1}$$

(ii) *Recognition and Significance:* After a pre-defined number of time steps (system's parameter), the network's states are matched to the closest pattern, i.e., we determine the index $i \in 1, ..., M$, such that

$$d_i = |m^\mu_{orig} - m^\mu_i| \text{ is minimal.}$$

If the output pattern matches the original one ($i \equiv orig$), we say that *recognition* occurs. Otherwise, we speak of *confusion* of the memories (concepts). The classification is considered *significant* only if

$$|d_i - d_j| < t \quad \forall j = 1, ..., M; \ (j \neq i),$$

where $t$ is a threshold parameter. With this scheme, we classify the memory recall into four groups and assign (numerical) labels.

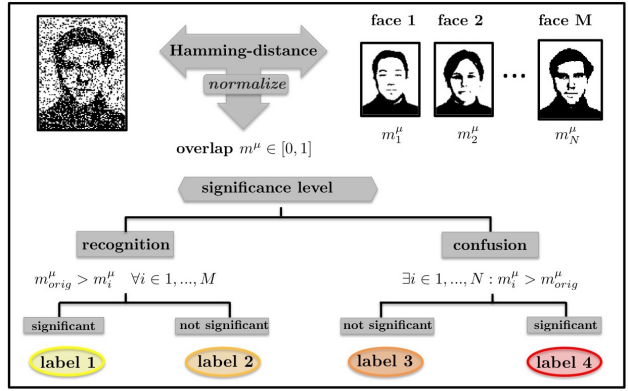

**Fig 6.** Calculation of recognition scores for measuring memory performance (see *Hopfield Recognition Toolbox*, current version available at GitHub: `https://github.com/MelWe/hopf-recognition`). We use the Hamming distance $m_i^\mu$ to measure the overlap between the current network state and the fixed points corresponding to known facial images. Confusion or recognition is characterized by $m_i^\mu$: if the overlap with the correct facial image is highest, we speak of *recognition*, otherwise of *confusion*. A threshold for the difference between the highest and second highest overlap determines whether the recognition or confusion was significant. According to this classification, we assign color labels to each trial which can be displayed in a heat map.

(iii) *Evaluation:* The recognition score was developed to evaluate the memory performance of our Hopfield neuronal network model over a broad range of injury and initial noise. For each pair of parameters $(inj, noise)$ we calculate the score as value of the significance label scaled by the accuracy of the recognition (overlap $m^\mu$).

The final result is a heat map (see Fig 2) that links recognition score, memory performance and noise handling to different levels of injury.

# Acknowledgments

MW was supported by a scholarship of the Konrad-Adenauer-foundation.